\documentclass[aip,jcp,amsmath,amssymb,longbibliography,floatfix,reprint]{revtex4-2}
\usepackage{graphicx}
\usepackage{dcolumn}
\usepackage{enumitem}
\usepackage{float}
\usepackage{bm}
\usepackage[mathlines]{lineno}
\usepackage[utf8]{inputenc}
\usepackage[T1]{fontenc}
\usepackage{etoolbox}
\usepackage{siunitx}
\usepackage{chemformula}

\newcommand{\csch}{\text{csch}}

\newcommand{\vecphi}{{\bm{\phi}}}
\newcommand{\vectheta}{{\bm{\theta}}}
\makeatletter
\def\@email#1#2{%
 \endgroup
 \patchcmd{\titleblock@produce}
  {\frontmatter@RRAPformat}
  {\frontmatter@RRAPformat{\produce@RRAP{*#1\href{mailto:#2}{#2}}}\frontmatter@RRAPformat}
  {}{}
}%
\makeatother
\begin{document}
\title{Pair Approximating the Action For Molecular Rotations in Path Integral Monte Carlo}
\author{Shaeer Moeed}
\author{Tobias Serwatka}
\author{Pierre-Nicholas Roy}%
 \email{pnroy@uwaterloo.ca.}
\affiliation{ 
Department of Chemistry, University of Waterloo, Waterloo, Ontario N2L 3G1, Canada
}%

\date{\today}

\begin{abstract}
    Typical path integral Monte Carlo approaches use the primitive approximation to compute the probability density for a given path. In this work, we develop the pair Discrete Variable Representation (pair-DVR) approach to study molecular rotations. The pair propagator, which was initially introduced to study superfluidity in condensed Helium, is naturally well-suited for systems interacting with a pair-wise potential. Consequently, paths sampled using the pair action tend to be closer to the exact paths (compared to primitive Trotter paths) for such systems leading to convergence with less imaginary time steps. Therefore, our approach relies on using the pair factorization approach in conjunction with a discretized Path Integral Ground State (PIGS) paradigm to study a chain of planar rotors interacting with a pair-wise dipole interaction. We first use the Wigner-Kirkwood density expansion to analyze the asymptotics of the pair propagator in imaginary time. Then, we exhibit the utility of the pair factorization scheme via convergence studies comparing the pair and primitive propagators. Finally, we compute energetic and structural properties of this system including the correlation function and Binder ratio as functions of the coupling strength to examine the behaviour of the pair-DVR method near criticality. Density Matrix Renormalization Group results are used for benchmarking throughout.
\end{abstract}

\maketitle

\section{Introduction}
Path integral Monte Carlo (PIMC) simulations are ubiquitous in the study of finite-temperature as well as ground state properties of many-body quantum systems.\cite{Barker,ceperley_review, blume, entanglement_entropy, pollock} Typically, the high-temperature propagator is approximated using the symmetric Trotter factorization which, while theoretically guaranteed to converge, can lead to long simulations times in the presence of large numbers of degrees of freedom.\cite{ceperley_review, higher_order_trotter} Moreover, path integral simulations aimed at numerically studying critical phenomena typically experience critical slowing down which causes the equilibration and auto-correlation times to become unfeasibly large.\cite{Wolff_critical_slow_down, topological_critical_slow_down}

The pair approximation is an alternative factorization scheme to Trotter factorization.\cite{Barker, pollock} It was initially studied in the context of superfluidity because it accounts for all $2$-body correlations exactly which improves PIMC convergence in imaginary time. It was shown numerically by Ceperley that paths sampled using the pair propagator are approximately $20$ times more accurate than the primitive factorization.\cite{ceperley_review} The pair approximation was also later used to compute real time correlation function for quantum fluids.\cite{makri} Finally, this factorization technique has been shown to be very useful for simulating systems with two-body zero range interactions.\cite{trapped_fermions_pair, fermi_gas_pair, zero_range_pair} However, all of these applications focused on factoring the many-body propagator in Cartesian coordinates.

Another problem that can be studied using PIMC is that of dipolar lattices. Such systems have recently garnered interest as potential platforms for quantum devices\cite{connor_entanglement_molecules_exp, Carr_2009} and are conventionally constructed using optical traps\cite{Rydberg_Optical_Lattice, dipolar_spin_exchange} or via chemical confinement of molecules with a permanent dipole moment such as water.\cite{exp_nanotube, endohedral_fullerites_exp} Such dipolar lattices are also good models for examining the ferroelectric properties of naturally occurring crystals such as Beryl and Cordierite because they have cavities capable of confining water molecules.\cite{gorshunov_2016_incipient} Recent work has suggested that these confined water molecules behave like planar rotors interacting via their electric dipole moments.\cite{belyanchikov_2020_dielectric}


In these confined molecular systems, at low temperatures, the vibrational degrees of freedom of each molecule are restricted to their respective ground states.
The orientational statistical mechanics of the many-body lattice is mostly determined by the rotational degrees of freedom of pinned molecules. 
Such effective Hamiltonians underscore the need to extend our PIMC capabilities to angular coordinates because those yield a natural parametrization of interacting rotational degrees of freedom. Recent Density Matrix Renormalization Group (DMRG) studies pertaining to $1$d lattices (chains) of dipolar planar rotors have shown that the system admits a quantum phase transition at $g \approx 0.5$.\cite{planar_rotor_QPT}

Previous work to improve PIMC efficiency for rotational motion focused on the development of a $4$th-order propagation scheme for linear rotors \cite{abolins2011ground, abolins2013erratum, whaley}. Here, we develop a pair product approximation for angular coordinates to study rotational motion because a pair propagation scheme will inherently account for all $2$-body correlations exactly with modest computational overhead. We use a $1$d lattice of dipolar planar rotors since that will not only enable us to benchmark the accuracy of our method using DMRG simulations but also allow us to examine the utility of the pair approximation in quantum critical regimes. Moreover, the techniques developed here can readily be extended to higher dimensional lattices to study the finite-temperature and ground state properties of crystals such as Beryl and Cordierite. The paper is organized as follows: We present  the theoretical formalism pertaining to the pair approximation in Sec. \ref{theory_section} and report our computational results in  Sec. \ref{results_section}. We conclude and discuss future research directions in Sec. \ref{conclusions_section}. 

\section{Theory} \label{theory_section}
\subsection{Planar Rotor Chain}
\begin{figure}[!b]
    \centering
    \includegraphics[width=0.9\linewidth]{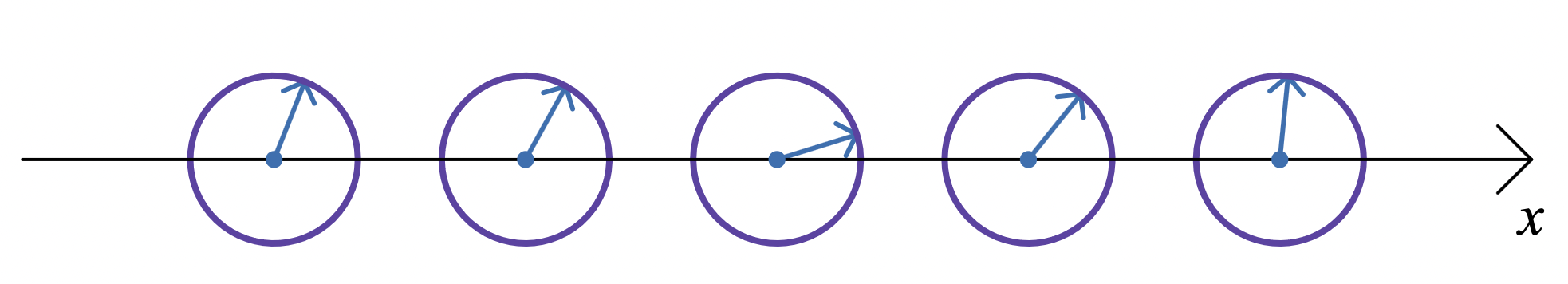}
    \caption{Schematic representation of a chain of planar rotors}
    \label{planar_rotors}
\end{figure}

The Hamiltonian for a linear chain of $N$ dipolar planar rotors is given by ($\hbar=1$),\cite{planar_rotor_QPT}
\begin{gather}
    H = \sum_{i=1}^{N} K_i + g\sum_{i=1}^{N-1} V_{i,i+1}
\end{gather}
A depiction of this model system is given in Fig. (\ref{planar_rotors}).
In angular position coordinates, the kinetic and potential energy operators take the form,
\begin{gather}
    K_i = l_i^2 = -\frac{\partial^2}{\partial \phi_i^2} \\
    V_{i,i+1} = (\sin(\phi_i)\sin(\phi_{i+1}) - 2 \cos(\phi_i) \cos(\phi_{i+1})) \label{two_body_potential}
\end{gather}
Here, $\phi_i$ represents the angular position of the $i$th rotor in the chain and $g$ is the coupling strength which is proportional to the cubic inverse distance between successive rotors ($r^{-3}$).  
This system undergoes a quantum phase transition at $g \approx 0.5$ which makes it a good model for optimizing rotational path integral ground state simulations in the many-body regime. In subsequent sections, we will develop the pair factorization approach in the context of this Hamiltonian. 
\subsection{The Pair Approximation}
The pair propagator can be derived using the Feynman-Kac theorem. Our construction follows the discussion provided in  Ceperley's review\cite{ceperley_review} closely. To this end, consider the low-temperature density operator $e^{-\beta H}$ expressed as a convolution of high temperature density operators, 
\begin{gather}
    \rho(\vecphi, \vecphi', \beta) = \langle \vecphi | e^{-\beta H} | \vecphi' \rangle \nonumber \\ = \int d\vecphi_1  ... d\vecphi_{P-1} \langle \vecphi | e^{-\tau H} | \vecphi_1 \rangle ... \langle \vecphi_{P-1} | e^{-\tau H} | \vecphi' \rangle
\end{gather}
where $\tau = \beta/P = 1/(PT)$ is the imaginary time and $T$ is the temperature. Note that in the above equation, we have inserted $P$ beads and $\vecphi_j = \left(\phi_j^1, ..., \phi_j^N\right)$ is the vector valued coordinate associated with the $j$th bead. For the remainder of this section, we will reserve superscripts for particle indices and subscripts for bead indices whenever we refer to the configuration variables (denoted $\vecphi$). We can express the high temperature density matrix as a continuum path integral,
\begin{equation} \label{path_integral}
    \rho(\vecphi, \vecphi', \tau) = \int D \vecphi_{\tau} \ e^{-\int_{0}^{\tau} d\tau' V(\vecphi(\tau'))}
\end{equation}
where the measure of the integral corresponds to $N$ independent Brownian motions on a circle. The path integral equation above is equivalent to the Feynman-Kac theorem, 
\begin{equation} \label{FK_theorem}
    \rho(\vecphi, \vecphi', \tau) = \rho_{0}(\vecphi, \vecphi', \tau) \left\langle e^{-\int_{0}^{\tau} d\tau' V(\vecphi(\tau'))} \right\rangle_{\mu(\vecphi, \vecphi')}
\end{equation}
where the angled brackets denote averaging over $N$ independent Brownian motions on a circle starting at $\vecphi$ and ending at $\vecphi'$. The transition probability density for this stochastic process (denoted $\mu(\vecphi, \vecphi')$) is simply the free propagator for a chain of $N$ planar rotor,\cite{S1_Brownian}
\begin{gather}
    \mu(\vecphi, \vecphi') = \langle \vecphi | e^{-\tau K} | \vecphi' \rangle = \prod_{j = 1}^{N} \langle \phi^{j} | e^{-\tau K_{j}} | \phi'^{j} \rangle \\ \langle \phi^{j} | e^{-\tau K_{j}} | \phi'^{j} \rangle = \frac{1}{2\pi}\sum_{m = -\infty}^{\infty} e^{-\tau m^2 + im(\phi^j - \phi'^j)}  \label{free_1_body_prop}
\end{gather}
Eq. \eqref{free_1_body_prop} is a Jacobi-Theta function. \cite{Whittaker_Watson_1996} In Eq. \eqref{FK_theorem}, note that we have simply restated the path integral in terms of statistics; the free propagator term represents the probability of taking any path between $\vecphi$ and $\vecphi'$ and the second term is the conditional expectation of the functional $\exp(-\int V(\tau') \ d\tau')$ over all such paths. Since our potential is pairwise additive,
\begin{gather}
    e^{\left(-\int_{0}^{\tau} d\tau' \ V(\vecphi(\tau'))\right)} = e^{\left(-\sum_{j=1}^{N-1} \int_{0}^{\tau} d\tau' \ V_{j,j+1}(\vecphi(\tau'))\right)} \nonumber \\ = \prod_{j = 1}^{N-1} e^{\left( - \int_{0}^{\tau} d\tau' \ V_{j,j+1}(\vecphi(\tau'))\right)}
\end{gather}
The pair approximation relies on ignoring all correlations that are higher than $2$-body in the averaging operation\cite{ceperley_review} of \eqref{FK_theorem},
\begin{gather}
    \left \langle \prod_{j = 1}^{N-1} e^{\left( - \int_{0}^{\tau} d\tau' \ V_{j,j+1}(\vecphi(\tau'))\right)} \right \rangle_{\mu(\vecphi, \vecphi')} \\ \approx \prod_{j = 1}^{N-1} \left \langle e^{\left(-\int_{0}^{\tau} d\tau' \ V_{j,j+1}(\vecphi(\tau'))\right)} \right \rangle_{\mu_j(\vecphi, \vecphi')}
\end{gather}
Here, the final averaging operation is taken to be over $2$ body random walks on a circle denoted $\mu_j$ where,
\begin{equation}
    \mu_{j}(\vecphi, \vecphi') = \langle \phi^{j} | e^{-\tau K_{j}} | \phi'^{j} \rangle \langle \phi^{j+1} | e^{-\tau K_{j+1}} | \phi'^{j+1} \rangle
\end{equation}
Now we can again use the Feynman-Kac theorem from Eq. \eqref{FK_theorem} to express the average over random walks as a product of two-body densities, 
\begin{gather}
    \prod_{j = 1}^{N-1} \left \langle e^{\left(-\int_{0}^{\tau} d\tau' \ V_{j,j+1}(\vecphi(\tau'))\right)} \right \rangle_{\mu_j(\vecphi, \vecphi')} \nonumber \\ = \prod_{j = 1}^{N-1}\frac{\rho^{(2)}_{j}(\vecphi, \vecphi', \tau)}{\rho^{(2)}_{0,j}(\vecphi, \vecphi', \tau)} 
\end{gather}
Here, $\rho^{(2)}_{j}$ is the two-body density operator,
\begin{gather}
    \rho^{(2)}_{j}(\vecphi, \vecphi', \tau) = \langle \phi^j, \phi^{j+1} | e^{-\tau H_{j}^{(2)}} | \phi'^{j}, \phi'^{j+1} \rangle 
    \\ H_{j}^{(2)} = K_j + K_{j+1} + V_{j, j+1} \label{2_body_H}
\end{gather}
Similarly, $\rho^{(2)}_{0,j}$ is the free two-body density operator, 
\begin{gather}
    \rho^{(2)}_{0,j}(\vecphi, \vecphi', \tau) = \langle \phi^j, \phi^{j+1} | e^{-\tau K_{j}^{(2)}} | \phi'^{j}, \phi'^{j+1} \rangle \\
    K_{j}^{(2)} = K_j + K_{j+1} \label{2_body_K}
\end{gather}
This finally yields the pair approximated form of the high-temperature density operator matrix elements,
\begin{equation} \label{pair_approx}
    \rho(\vecphi, \vecphi', \tau) \approx \rho_{0}(\vecphi, \vecphi', \tau) \prod_{j=1}^{N-1} \frac{\rho^{(2)}_{j}(\vecphi, \vecphi', \tau)}{\rho^{(2)}_{0,j}(\vecphi, \vecphi', \tau)} 
\end{equation}
The pair approximation is mainly motivated by the form of the pairwise nature of the potential. The two-body correlations are entirely accounted for by construction. However, many-body correlations which are often less relevant are ignored. It would be interesting to explore the utility of the pair approximation in the presence of many-body interactions. 
\subsection{Pair Density Error} \label{Pair_Density_Error}
Usual approaches for path integral simulations employ the symmetric Trotter approximation, 
\begin{equation}
    \langle \vecphi | e^{-\tau H} | \vecphi  \rangle \approx e^{-\frac{\tau}{2} (V(\vecphi) + V(\vecphi'))} \langle \vecphi | e^{-\tau K} | \vecphi  \rangle
\end{equation}
for which it is easy to show that the error is $\mathcal{O}(\tau^3)$ using the 
Baker–Campbell–Hausdorff (BCH) formula. To compare the pair density with the symmetric Trotter factorization, it is instructive to derive the density error in the pair approximation. However, since the pair density is defined in terms of matrix elements, it is convenient to use the asymptotic Wigner-Kirkwood series for this purpose.\cite{WK_expansion, WK_expansion2, makri_WK_exp} To this end, note first that the density operator satisfies the Bloch equation,
\begin{gather}
    \left(\frac{\partial}{\partial \tau} + H\right) \rho(\vecphi, \vecphi', \tau) = 0 \\
    \lim_{\tau \rightarrow 0} \rho(\vecphi, \vecphi', \tau) = \delta(\vecphi - \vecphi')
\end{gather}
Since we want a series expansion for the quotient of the full propagator and the free propagator, we assume that the solution takes the form, 
\begin{equation}
    \rho(\vecphi, \vecphi', \tau) = \rho_0 (\vecphi, \vecphi', \tau) F(\vecphi, \vecphi', \tau)
\end{equation}
Then, using the Bloch equation for the free propagator as well as the full propagator, we get the following differential equation for $F$ (see Supplementary Material, Sec. A for details\cite{suppl}),
\begin{equation} \label{F_PDE}
    \left(\frac{\partial}{\partial \tau} - \Delta_{\vecphi} - 2 \frac{\nabla_{\vecphi} \rho_0}{\rho_0} \cdot \nabla_{\vecphi} + V\right) F = 0
\end{equation}
where we have used the notation,
\begin{equation}
    \nabla_{\vecphi} = \left(\frac{\partial}{\partial \phi_1}, ..., \frac{\partial}{\partial \phi_N} \right), \ \ \Delta_\vecphi = \sum_{i=1}^{N} \frac{\partial^2}{\partial \phi_i^2}
\end{equation}
It can be shown using properties of Jacobi-Theta functions that to first order in $\tau$ (see Supplementary Material, Sec. B for details),
\begin{equation}
    \left(\frac{\nabla_{\vecphi} \rho_0}{\rho_0}\right) \approx -\frac{1}{2\tau} (\vecphi - \vecphi') \label{d_prop_over_prop_approx}
\end{equation}
To understand this approximation in more detail, note that the free propagator can be written as follows using the Poisson summation formula,\cite{pinsky_2023_introduction}
\begin{equation}
    \rho_0^{(1)}(\phi^j, \phi'^j, \tau) = \frac{1}{\sqrt{4\pi \tau}}\sum_{k=0}^{\infty} e^{-(\phi^j - \phi'^j - 2\pi k)^2/4\tau} 
\end{equation}
This is a sum over Gaussians each with variance $2\tau$ and mean $2\pi k$ for $k\in \mathbb{Z}$. When $\tau << 2\pi$, each of these Gaussians are localized to their respective $2\pi$ intervals which yields Eq. \eqref{d_prop_over_prop_approx} for $\left(\frac{\nabla_{\vecphi} \rho_0}{\rho_0}\right)_{j}$ on the interval $[-\pi, \pi)$ again. Therefore, this approximation is valid whenever $\tau$ is small enough such that these Gaussians are sufficiently peaked in their $2\pi$ intervals that they do not appreciably affect the free propagator in other $2\pi$ intervals. Since in our path integral simulations, we expand the high temperature propagator using the pair factorization, we can use this to analyze the $\tau$ error in our approximation. 

Using this, we get the same differential equation for $F$ as in the case of Cartesian coordinates for linear motion,\cite{WK_expansion2}
\begin{equation}
    \left(\frac{\partial}{\partial \tau} - \Delta_{\vecphi} + \frac{1}{\tau} (\vecphi - \vecphi') \cdot \nabla_{\vecphi} + V\right) F = 0 \label{diff_eqn_F}
\end{equation}
Now, we can expand $F$ as a power series in $\tau$,
\begin{equation} \label{F_expansion}
    F = \sum_{n=0}^{\infty} \frac{(-\tau)^n}{n!} P_n(\vecphi, \vecphi')
\end{equation} 
Note that we can set $P_0(\vecphi, \vecphi') = F(\vecphi, \vecphi', 0) = 1$ because as $\tau \rightarrow 0$, the system becomes classical and $F \rightarrow e^{-\tau V} \approx 1 - \tau V$. This differential equation can be solved using Perelomov's recursion relation,\cite{WK_expansion2}
\begin{gather} \label{Perelomov_recursion}
    P_n(\vecphi, \vecphi') = n \int_0^1 dt \ t^{n-1} V(\vectheta(t)) P_{n-1}(\vectheta(t), \vecphi') \nonumber \\ - n \int_0^1 dt \ t^{n-1} \Delta_\vectheta P_{n-1}(\vectheta(t), \vecphi') 
\end{gather}
where $\vectheta(t) = t\vecphi + (1-t)\vecphi'$. Since $P_0 = 1$, $P_1$ can be readily computed using the above equation,
\begin{gather} \label{P_1_eqn}
    P_1(\vecphi, \vecphi') = \int_{0}^{1} V(\vectheta(t)) \ dt = \sum_{i=1}^{N-1} W_{i, i+1} (\vecphi, \vecphi') \\
    W_{i, i+1} (\vecphi, \vecphi') = \int_0^1 V_{i, i+1}(\vectheta(t))
\end{gather}
where we have used the fact that for us, the potential is pairwise additive. For $P_2$, we can now use $P_1$ and integrate the first term by parts to get (see Supplementary Material, Sec. C for details),
\begin{gather}
    P_2(\vecphi, \vecphi') = \sum_{j=1}^{N-1} W_{j,j+1}^2 + 2\sum_{i<j}^{N-1} W_{i,i+1}W_{j,j+1}  \nonumber \\
    - 2\sum_{i=1}^{N-1} \int_0^1 dt \ t \left(\frac{\partial^2 W_{i, i+1}}{\partial \theta_i^2} + \frac{\partial^2 W_{i, i+1}}{\partial \theta_{i+1}^2}\right) \label{P_2_eqn_1} 
\end{gather}
where $W_{j,j+1}=W_{j,j+1}(\vectheta(t), \vecphi')$ in the above equation. Now, we can return to our original goal of comparing the two sides of Eq. \eqref{pair_approx} in powers of $\tau$. For the right side, we have products over pair densities. Expanding each using Eq. \eqref{F_expansion} and multiplying gives (to third order),
\begin{gather}
    \prod_{n=1}^{N-1} F^{(2)}_n \approx 1 - \tau Q_1 + \frac{\tau^2}{2} Q_2 - \frac{\tau^3}{6} Q_3 \label{pair_density_expansion} 
\end{gather}
To compare the first order terms, note that,
\begin{gather}
    Q_1 = \sum_{n=1}^{N-1} P_{1,n}^{(2)}
\end{gather} 
From Eq. \eqref{P_1_eqn}, it can be seen that for the $2$-body problem, we get $P_{1,n}^{(2)} = W_{n,n+1}$ which implies that the series' in Eqs. $\eqref{F_expansion}$ and \eqref{pair_density_expansion} agree to first order. For the second order term, note that $Q_2$ takes the form,
\begin{gather}
    Q_2 = \sum_{n=1}^{N-1}P_{2,n}^{(2)} + 2\sum_{n<m}^{N-1}P_{1,n}^{(2)}P_{1,m}^{(2)} \label{Q_2}
\end{gather}
As before, we can determine $P_{2,n}^{(2)}$ from Eq. \eqref{P_2_eqn_1}. Substituting this and $P_{1,n}^{(2)}$ into Eq. $\eqref{Q_2}$
gives the right side of $\eqref{P_2_eqn_1}$ which shows that the pair density is exact to second order. We can also use this approach to show that the error in the pair approximation is $\mathcal{O}(\tau^3)$. To see this, it is sufficient to compare the diagonal terms ($\vecphi=\vecphi'$) of the Wigner-Kirkwood expansion. These can be computed by taking the limit $\vecphi \rightarrow \vecphi'$. In particular for $P_3$, we get,\cite{WK_expansion2}
\begin{gather}
    P_3 = V^3 - V \Delta_{\vecphi} V - \frac{1}{2} (\nabla_\vecphi V)^2 + \frac{1}{10} \Delta_{\vecphi}^2 V \label{P_3_diag_eqn}
\end{gather}
To compare $P_3(\vecphi, \vecphi)$ to $Q_3(\vecphi, \vecphi)$, we need to express $Q_3$ using products and sums of $2$-body terms as before,
\begin{gather}
    Q_3 = \sum_{n=1}^{N-1} P_{3,n}^{(2)} + 6\sum_{n<m<l}P_{1,n}^{(2)}P_{1,m}^{(2)}P_{1,l}^{(2)} \nonumber \\ + \ 3\sum_{n<m}^{N-1} (P_{2,n}^{(2)}P_{1,m}^{(2)} + P_{1,n}^{(2)}P_{2,m}^{(2)}) \label{Q_3_eqn}
\end{gather}
Using Eqs. \eqref{P_3_diag_eqn} and \eqref{Q_3_eqn}, we find that $P_3$ and $Q_3$ agree for all terms in Eq. $\eqref{P_3_diag_eqn}$ except for $(\nabla_{\vecphi} V)^2/2$ (see Supplementary Material, Sec. D for details). The exact difference between $P_3$ and $Q_3$ takes the form,
\begin{equation}
    P_3 - Q_3 = -\sum_{n=2}^{N-1} (\nabla V_{n-1,n})_{n} (\nabla V_{n,n+1})_{n} \neq 0 \label{P_3_error_term}
\end{equation}
Therefore, the error in the pair density is $\mathcal{O}(\tau^3)$. It is also instructive to note that the error function in Eq. $\eqref{P_3_error_term}$ is a sum of terms that correspond to $3$-body correlations. This is unsurprising since we derived the pair approximation by neglecting all higher than $2$-body correlations. Moreover, in the higher order terms in $\tau$ the pair factorization again accounts for $2$-body correlations exactly. Consequently, for those terms as well, we expect that the error would arise due to higher than $2$-body correlations. This means that the pair factorization leads to much smaller error coefficients (in powers of $\tau$) when $2$-body correlations dominate the behavior of a system. Therefore, for a given $\tau$, we see far less systematic error in such regimes compared to the symmetric Trotter factorization even though the latter also has error $\mathcal{O}(\tau^3)$. In regimes where many-body correlations become significant (such as near criticality), the pair factorization error increases. 
\subsection{Path Integral Ground State  Estimators}
We will use the pair approximation to compute ground state properties via Path Integral Ground State (PIGS). In this paradigm, ground state operator expectation values are computed as,\cite{blume} 
\begin{equation} \label{PIGS_eqn}
    \langle O \rangle = \frac{1}{Z}\langle \Psi_T | e^{-\frac{\beta H}{2}} O  e^{-\frac{\beta H}{2}} | \Psi_T \rangle
\end{equation}
where $Z$ is the pseudo-partition function, 
\begin{equation}
    Z = \langle \Psi_T | e^{-\beta H} | \Psi_T \rangle
\end{equation}
For this work, we will restrict to the constant trial wave-function in the position representation ($\langle \vecphi | \Psi_T\rangle = 1$). The PIGS ground state energy estimator is given by,\cite{blume,zhang2024path}
\begin{gather} \label{energy_pigs_estimator}
    E_{0} =
     A \int d \vecphi_0... d\vecphi_{P} \prod_{i=0}^{P-1} \rho (\vecphi_{i}, \vecphi_{i+1}, \tau)\ V(\vecphi_{P})
\end{gather}
where $A = 1/2\pi Z$. Expanding $\rho(\tau)$ using the pair approximation from Eq. \eqref{pair_approx}, we get,
\begin{gather}
    E_{0} =
    A \int d \vecphi_0... d\vecphi_P \ \prod_{i=0}^{P-1} \rho_0 (\vecphi_{i}, \vecphi_{i+1}, \tau) \nonumber \\ \prod_{j=1}^{N-1} \frac{\rho^{(2)}_{j}(\phi_{i}^{j}, \phi_{i}^{j+1},\phi_{i+1}^{j},\phi_{i+1}^{j+1}, \tau)}{\rho^{(2)}_{0,j}(\phi_{i}^{j}, \phi_{i}^{j+1},\phi_{i+1}^{j},\phi_{i+1}^{j+1}, \tau)}\ V(\vecphi_{P})
\end{gather}
In addition to the energy, we also want to study structural properties. Since these are diagonal in the position representation, they can be evaluated using Eq. \eqref{PIGS_eqn}. In particular, we need to compute the angular correlation,
\begin{equation} \label{corr_def}
    C = \sum_{i=1}^{N-1} \cos(\phi^{i} - \phi^{i+1})
\end{equation}
The PIGS estimator for this is given by, 
\begin{gather} 
    C_0 =
    A \int d \vecphi_0... d\vecphi_P \prod_{i=0}^{P-1} \rho_0 (\vecphi_{i}, \vecphi_{i+1}, \tau) \nonumber \\ \prod_{j=1}^{N-1} \frac{\rho^{(2)}_{j}(\phi_{i}^{j}, \phi_{i}^{j+1},\phi_{i+1}^{j},\phi_{i+1}^{j+1}, \tau)}{\rho^{(2)}_{0,j}(\phi_{i}^{j}, \phi_{i}^{j+1},\phi_{i+1}^{j},\phi_{i+1}^{j+1}, \tau)} \ C(\bm{\phi}_{P/2 + 1}) \label{corr_pigs_estimator}
\end{gather}
Finally, we also need to compute the Binder ratio defined as,\cite{Binder_ratio_finite_size_scaling, planar_rotor_QPT} 
\begin{equation} \label{Binder_Def}
    U = 1 - \frac{\langle M_x^4 \rangle}{3 \langle M_x^2 \rangle^2}
\end{equation}
where $M_x = \sum_{i=1}^{N} \cos(\phi_i)$ is the net polarization across the chain. The second and fourth powers of the polarization are given by,
\begin{gather}
    M_x^2 = \sum_{i,j=1}^{N} \cos(\phi_i)\cos(\phi_j) \\
    M_x^4 = \sum_{i,j,k,l=1}^{N} \cos(\phi_i)\cos(\phi_j)\cos(\phi_k)\cos(\phi_l)
\end{gather}
$M_x^k$ for $k \in \{2,4\}$ is diagonal in the position representation so it can be computed using an estimator analogous to \eqref{corr_pigs_estimator}, 
\begin{gather} 
    M_{x,0}^k = A \int d \vecphi_0... d\vecphi_P \prod_{i=0}^{P-1} \rho_0 (\vecphi_{i}, \vecphi_{i+1}, \tau) \nonumber \\ \prod_{j=1}^{N-1} \frac{\rho^{(2)}_{j}(\phi_{i}^{j}, \phi_{i}^{j+1},\phi_{i+1}^{j},\phi_{i+1}^{j+1}, \tau)}{\rho^{(2)}_{0,j}(\phi_{i}^{j}, \phi_{i}^{j+1},\phi_{i+1}^{j},\phi_{i+1}^{j+1}, \tau)} \ M_x^k(\bm{\phi}_{P/2 + 1}) \label{binder_pigs_estimator}
\end{gather} 

\subsection{Discrete Variable Representation} \label{DVR_section}
To optimize our path integral simulations, we use the Colbert-Miller\cite{dvr} Discrete Variable Representation (DVR)\cite{light1985generalized,light2000dvr} to discretize the continuous angular degrees of freedom. In this discretization scheme, the grid points for each $\phi$ are linearly spaced, 
\begin{equation} \label{phi_grid}
    \{\phi: \phi \in [0, 2\pi)\} \rightarrow \{2\pi j/M : j\in 0,...,M-1\} 
\end{equation}
where $M = 2l+1$ and $l$ is the number of kinetic energy eigenstates used to construct the DVR representation. The $1$-body kinetic energy operator is given by the matrix,\cite{dvr}
\begin{gather} \label{DVR_K}
    K^{(1)}_{i, j} = \langle 2\pi i/M|K^{(1)}|2\pi j/M \rangle \nonumber \\
    = \frac{d(d+1)}{3}, \ \text{for} \ i = j \\
    = (-1)^{i-j} \frac{\cos(\pi (i-j)/M)}{2\sin^2(\pi (i-j)/M)},  \ \text{for} \ i \neq j 
\end{gather}
The $2$-body potential operator is diagonal in this representation,
\begin{gather}
    V^{(2)}_{ijkl} = \langle 2\pi i/M, 2 \pi j/M |V^{(2)}|2\pi k/M, 2 \pi l/M\rangle \nonumber \\ = V^{(2)}(2\pi i/M, 2 \pi k/M) \delta_{i,j}\delta_{jl}
\end{gather}
where $V^{(2)}$ here is given by the Eq. \eqref{two_body_potential}. We will use this discretization scheme to cast our path integrals into path sums.\cite{zhang2024path} 

To do this, first note that the required 2-body Hamiltonian $H^{(2)}$ (Eq. \eqref{2_body_H})  can be computed as a matrix using the DVR representations of $K$ and $V$ provided above. Then, the corresponding 2-body propagator required for the pair approximation can be numerically tabulated in the DVR position representation. Similarly, we can diagonalize the matrix associated with $K_i$ in the DVR representation to compute the $1$-body free propagator. To use the pair density in path integral simulations, we need to tabulate the following two tensors, 
\begin{gather}
    A_{ij}^{kl} = \frac{\langle 2\pi i/M, 2\pi j/M | e^{-\tau H^{(2)}} | 2\pi k/M, 2\pi l/M \rangle}{\langle 2\pi i/M, 2\pi j/M | e^{-\tau K^{(2)}} | 2\pi k/M, 2\pi l/M \rangle} \label{effective_potential_pair_dvr} \\
    B_i^j = \langle 2\pi i/M | e^{-\tau K^{(1)}} | 2\pi j/M \rangle
\end{gather}
$B$ allows us to sample the free propagator part in \eqref{pair_approx} and $A$ is stored to sample the $2$-body effective potential part in \eqref{pair_approx}. Note that $B$ is a rank 2 tensor which can be stored as a table as specified. To store $A$ as a matrix, we can use lexicographic ordering to get,
\begin{gather}
    A_i^j = A_{i_1,i_2}^{j_1,j_2}
    \\ i_1 = \lfloor i/M \rfloor, \ i_2 = i - i_1 M
    \\ j_1 = \lfloor j/M \rfloor, \ j_2 = j - j_1 M
\end{gather}
Both $A_{i}^{j}$ and $B_{i}^{j}$ are computed and tabulated before the Monte Carlo run to minimize on-the-fly calculations. 
\subsection{Sequentially Gibbs Sampling The Pair Density}
To estimate the ground state properties specified by Eqs. \eqref{energy_pigs_estimator}, \eqref{corr_pigs_estimator} and \eqref{binder_pigs_estimator}, we use Markov chain Monte Carlo integration \cite{mcmc_algo} with a sequential Gibbs sampler.\cite{gibbs} Consider the pair factorization probability density,
\begin{equation}
    P(\Omega, \tau) = \prod_{i,j} \rho_0 (\vecphi_{i}, \vecphi_{i+1}, \tau) \frac{\rho^{(2)}_{j}(\vecphi_{i}, \vecphi_{i+1}, \tau)}{\rho^{(2)}_{0,j}(\vecphi_{i}, \vecphi_{i+1}, \tau)}
\end{equation}
Here, $i$ is the bead index, $j$ is the particle index and $\Omega=(\vecphi_1, ..., \vecphi_P)$. In Gibbs sampling, the system moves are determined using conditional probabilities on a discrete state space. We can use our pre-computed probability tables $A_{ijkl}$ and $B_{ij}$ to determine the conditional probabilities for the $p$th bead and $n$th rotor coordinate as follows,
\begin{gather}
    P(\phi_p^n(t+1)=2\pi j/M| \Phi(t,t+1)) = \frac{P_B P_A}{P_M} \label{cond_prob_gibbs}
\end{gather}
where, $t$ is the Monte Carlo simulation time-step and $P_A, P_B$ and $P_M$ are given by,
\begin{gather}
    P_B = B_{i_2}^{j}B_{j}^{i_8} \label{P_B_gibbs}\\
    P_A = A_{i_1,i_2}^{i_4,j}A_{i_2,i_3}^{j,i_6}A_{i_4,j}^{i_7,i_8}A_{j,i_6}^{i_8,i_9} \label{P_A_gibbs}\\
    P_M = {\sum_{k} A_{i_1,i_2}^{i_4,k}A_{i_2,i_3}^{k,i_6}A_{k,i_6}^{i_8,i_9} B_{i_2}^{k}B_{k}^{i_8}} \label{P_M_gibbs}
\end{gather}
The variable being conditioned on in Eq. \eqref{cond_prob_gibbs} is the configuration vector $\Phi(t,t+1)$,
\begin{gather}
    \Phi(t,t+1) = (\vecphi_0(t+1),...,\tilde{\vecphi}_p(t,t+1),...,\vecphi_P^N(t)) \label{var_cond_gibbs}
\end{gather}
Here, the components of the vector in Eq. \eqref{var_cond_gibbs} take the form,
\begin{gather}
    \vecphi_k(s) = (\phi_k^1(s),...,\phi_k^N(s)), \ 0 \leq k \leq P, k\neq p \\
    \tilde{\vecphi}_p(t,t+1) = (\phi_p^1(t+1),...,\phi_p^{n-1}(t+1), \nonumber \\ \phi_p^{n+1}(t), ... , \phi_p^N(t))
\end{gather}
The indices $i_l$ for $l \in \{1,2,3, 7, 8, 9\}$  in Eqs. \eqref{P_B_gibbs}, \eqref{P_A_gibbs} and \eqref{P_M_gibbs} are defined by,
\begin{gather}
    \phi^{n-1+l}_{p-1}(t) = \frac{2\pi i_l}{M}, \ \ l \in \{1,2,3\} \\
    \phi^{n-1+l}_{p+1}(t+1) = \frac{2\pi i_l}{M}, \ \ l \in \{7,8,9\}
\end{gather}
Similarly, indices $i_4$ and $i_6$ are defined by,
\begin{equation}
    \phi^{n-1}_{p}(t+1) = \frac{2\pi i_4}{M}, \ \ \phi^{n+1}_{p}(t) = \frac{2\pi i_6}{M}
\end{equation}
Eqs. \eqref{P_B_gibbs}, \eqref{P_A_gibbs} and \eqref{P_M_gibbs} above correspond to $n,p$ such that $0<p<P$, $1<n<N$. For the terminal bead and rotor indices, at least one of $p \pm 1$ or $n \pm 1$ would be negative. Therefore, terms containing $i_l$ for such an $l$ would simply be excluded. In the sequential Gibbs sampling algorithm, we iterate over all the beads and all the rotors in each step.\cite{gibbs} For each of the $N(P+1)$ variables, we then compute the conditional probabilities of the $M$ possible moves (Eq. \eqref{cond_prob_gibbs}) and then uniformly sample the distribution associated with said probabilities.\cite{geman_1984_stochastic}
\begin{figure}
  \centering
\includegraphics[width=\columnwidth]{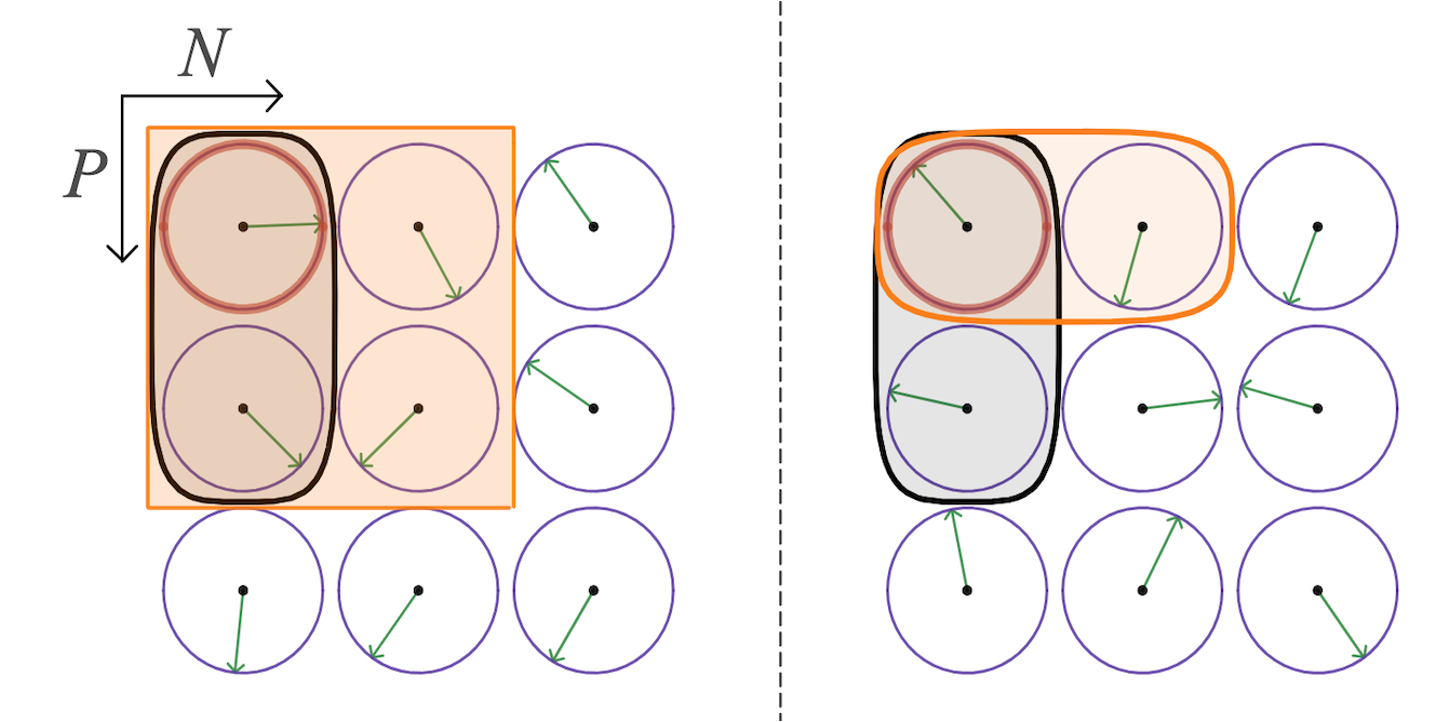}
\caption{Comparison of sampling probability calculations between the pair and symmetric Trotter factorization in the context of Gibbs sampling. For simplicity, we compare the $p=1,n=1$ variable in $NP$-space. The left panel corresponds to the pair factorization while the right panel exhibits the case for symmetric Trotter factorization. The top-left $4$ shaded rotors in the left panel represent $A_{j,i_6}^{i_7,i_8}$ whereas the grey vertical region encapsulating the first rotor represents $B_j^{i_7}$. $B_j^{i_7}$ is also present in the Trotter case since it represents the free particle density. However, for symmetric Trotter, we have a $2$-body term: $e^{-\tau V_{j,i_6}}$ instead of the $4$-body $A_{j,i_6}^{i_7,i_8}$ term needed for the pair approximation case.}
\label{sampling_comparison}
\end{figure}

Figure \ref{sampling_comparison} compares the sampling approaches for the pair and symmetric Trotter factorization in the context of Gibbs sampling. For the pair factorization, to calculate each transition probability, we need to account for the configurations of all the surrounding particles in $NP$-space. In this context, the symmetric Trotter propagator is simpler since the transition probability only depends on the $(p,n \pm 1)$ and $(p \pm 1,n)$ particle configurations. Note that in the figure, we exhibit the case for $(p,n) = (0,1)$ for simplicity. For non-terminal beads and rotors, analogous terms for other surrounding variables would also be present as in Eqs. \eqref{P_B_gibbs}, \eqref{P_A_gibbs} and \eqref{P_M_gibbs}. 
The advantage of Gibbs sampling is that it is rejection-free. Moreover, using an optimized grid such as the Colbert-Miller DVR means that $M$ is small which also optimizes the simulation performance. In particular, for this problem, $M=5$ is sufficient to converge to the exact results.\cite{zhang2024path} 
\subsection{The Sign Problem}
It is worth noting that using the DVR representation for computing propagators can lead to negative density matrix elements when $\tau$ is very small (which happens when the number of beads being used is large). This can be seen by considering the first order Taylor expansion of the $1$-body free propagator in $\tau$,
\begin{equation} \label{taylor_DVR_K}
    \langle 2\pi i/M | e^{-\tau K^{(1)}} | 2 \pi j/M \rangle \approx \delta_{ij} - \tau K^{(1)}_{ij}
\end{equation}
Since $K_{ij}^{(1)}$ is non-stoquastic (has positive off-diagonal matrix elements), the off-diagonal matrix elements of \eqref{taylor_DVR_K} can be negative for small $\tau$. This is problematic because density matrix elements are used to assign probabilities to different configurations in a Monte Carlo simulation which makes negative definite densities unusable in this context. It can also be shown that when $\tau$ is large, there is no sign problem in the free propagator. Moreover, the critical value of $\tau$ such that this issue arises in the free propagator depends on the number of grid points $M$. It was computationally shown in Ref. \onlinecite{zhang2024path} that for $M=5$, the critical $\tau$ value is approximately $0.2$. 

The same argument as that used for the free propagator shows that for small enough $\tau$, the $2$-body effective potential part of the pair propagator given in Eq. \eqref{effective_potential_pair_dvr} can have negative density matrix elements. However, unlike the free propagator, this can not be circumvented by increasing the number of grid points. For our simulations, when we encounter the sign problem with the pair propagator, we keep track of the overall sign of Eq. \eqref{pair_approx} during the simulation and multiply the estimators by that sign.\cite{sign_prob} While we can use this approach for mild sign problems such as the one we encounter in our simulations, it should be noted that if the sign problem is severe, this approach can make the integrand oscillatory and lead to an exponential slowdown in Monte Carlo convergence. 
\subsection{Numerical Matrix Multiplication}
For small systems, we test our approach using Numerical Matrix Multiplication (NMM).\cite{nmm_ref} This involves casting our path integral (Eq. \eqref{PIGS_eqn}) as a matrix multiplication problem,
\begin{equation}
    \langle f \rangle = \frac{1}{Z} \left(\prod_{j=1}^{P/2} \hat{\rho} \right) \hat{f} \left(\prod_{j=1}^{P/2} \hat{\rho} \right)
\end{equation}
where $\hat{\rho}$ is the high temperature density matrix in the position representation and $\hat{f}$ represents the matrix associated with the property $f$ in the position representation. The product in the above equation is matrix multiplication. For our work, we construct the pair density for $3$ rotors as follows,
\begin{equation}
    \rho_{ijk}^{lmn} = B_i^l B_j^m B_k^n A_{ij}^{lm} A_{jk}^{mn}
\end{equation}
where $B$ and $A$ are the matrices constructed in Sec. \ref{DVR_section}. $\rho_{ijk}^{lmn}$ can be represented as a matrix using lexicographic ordering, 
\begin{gather}
    \hat{\rho}_{rs} = \rho_{ijk}^{lmn} \\
    r = iM^2+jM+k, \ \ s = lM^2 + mM + n
\end{gather}
\section{Results \& Discussion} \label{results_section}
\subsection{Convergence Studies} \label{converge}
\begin{figure}
  \centering
\includegraphics[width=\columnwidth]{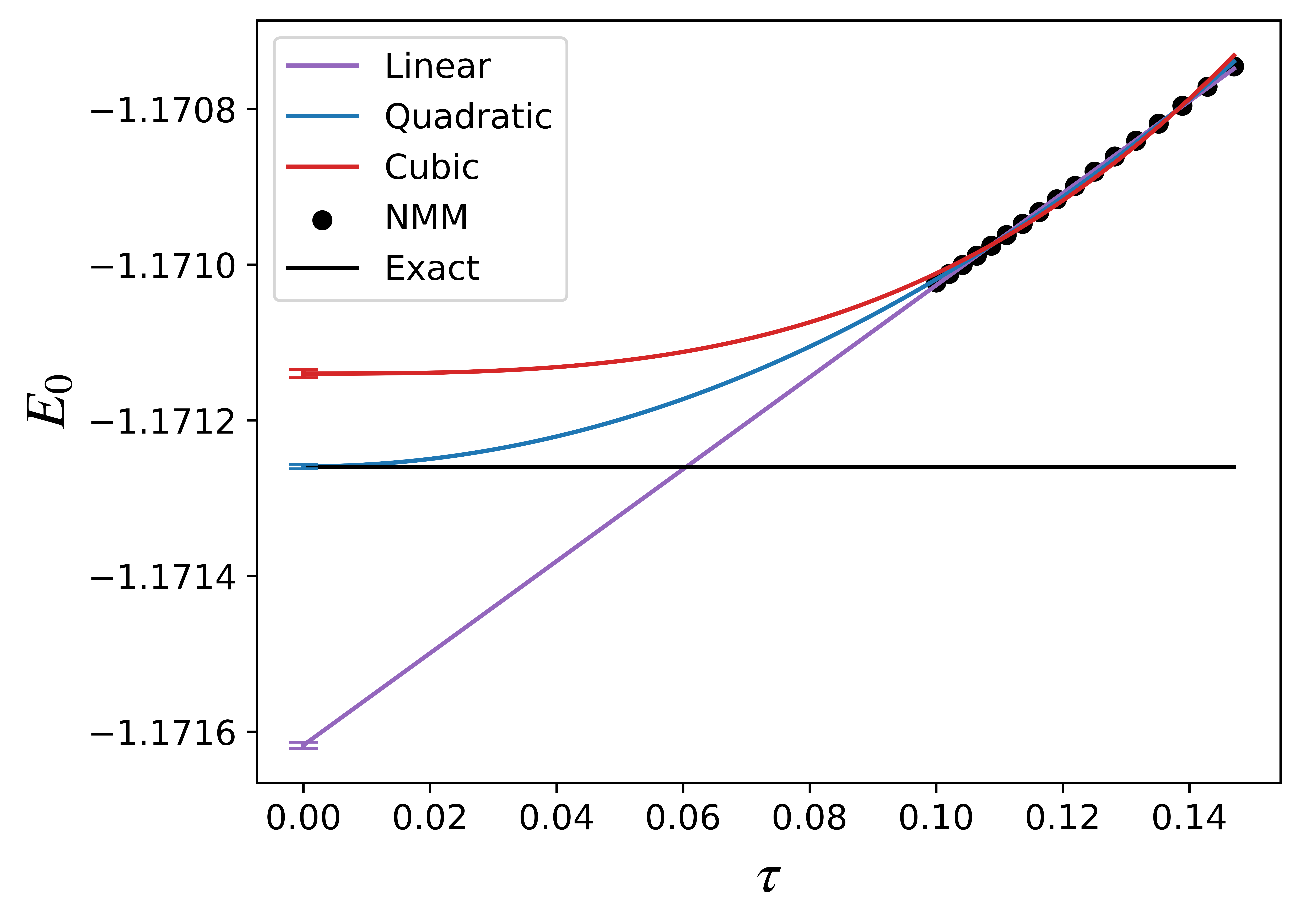}
\caption{NMM fit comparison results with the pair propagator for $N=3$, $g=1.0$ and $M=5$. Exact ground state energy results are plotted for comparison.} 
\label{Pair_Convergence}
\end{figure}

We first determine the PIGS ground state energy estimator extrapolation function in the limit of $\tau \rightarrow 0$ computationally. This is done by fitting the pair propagated NMM energy results for $66 \leq P \leq 100$. As for all other results in this work, we use $\beta = 10.0$ since that has been shown to be sufficient for PIGS convergence in past work.\cite{zhang2024path} Figure \ref{Pair_Convergence} show the convergence fits corresponding to $N = 3$ with $M = 5$. We compare three different fitting functions: linear, quadratic ($E_0(\tau) = E_0 + \tau^2 b$) and cubic ($E_0(\tau) = E_0 + \tau^3 b$).  

The error bars at $\tau = 0$ represent the fitting errors for each of the extrapolations. The quadratic function fits the NMM data and converges to the correct energy. This is unsurprising since the density error (as in the case of the symmetric Trotter approximation) is $O(\tau^3)$ and we know from previous work that the asymptotic behaviour in the limit of $\tau \rightarrow 0$ is well described by quadratic fitting functions for the symmetric Trotter factorization. The exact ground state energy for this result is calculated by diagonalizing the $3$-body Hamiltonian to calculate the corresponding PIGS estimate using Eq. \eqref{PIGS_eqn},
\begin{gather}
    \langle E_0 \rangle = \frac{1}{Z}\sum_{i,j,k} \langle \vecphi_i|E_k\rangle \langle E_k|\vecphi_j\rangle E_k e^{-\beta E_k} \\ Z = \sum_{i,j,k} \langle \vecphi_i|E_k\rangle \langle E_k|\vecphi_j\rangle e^{-\beta E_k}
\end{gather}
This is done instead of using $E_0$ obtained by diagonalizaing $H$ so as to include the effects of using a finite $\beta$ in the fit comparison. 

\begin{figure} [h]
  \centering
\includegraphics[width=\columnwidth]{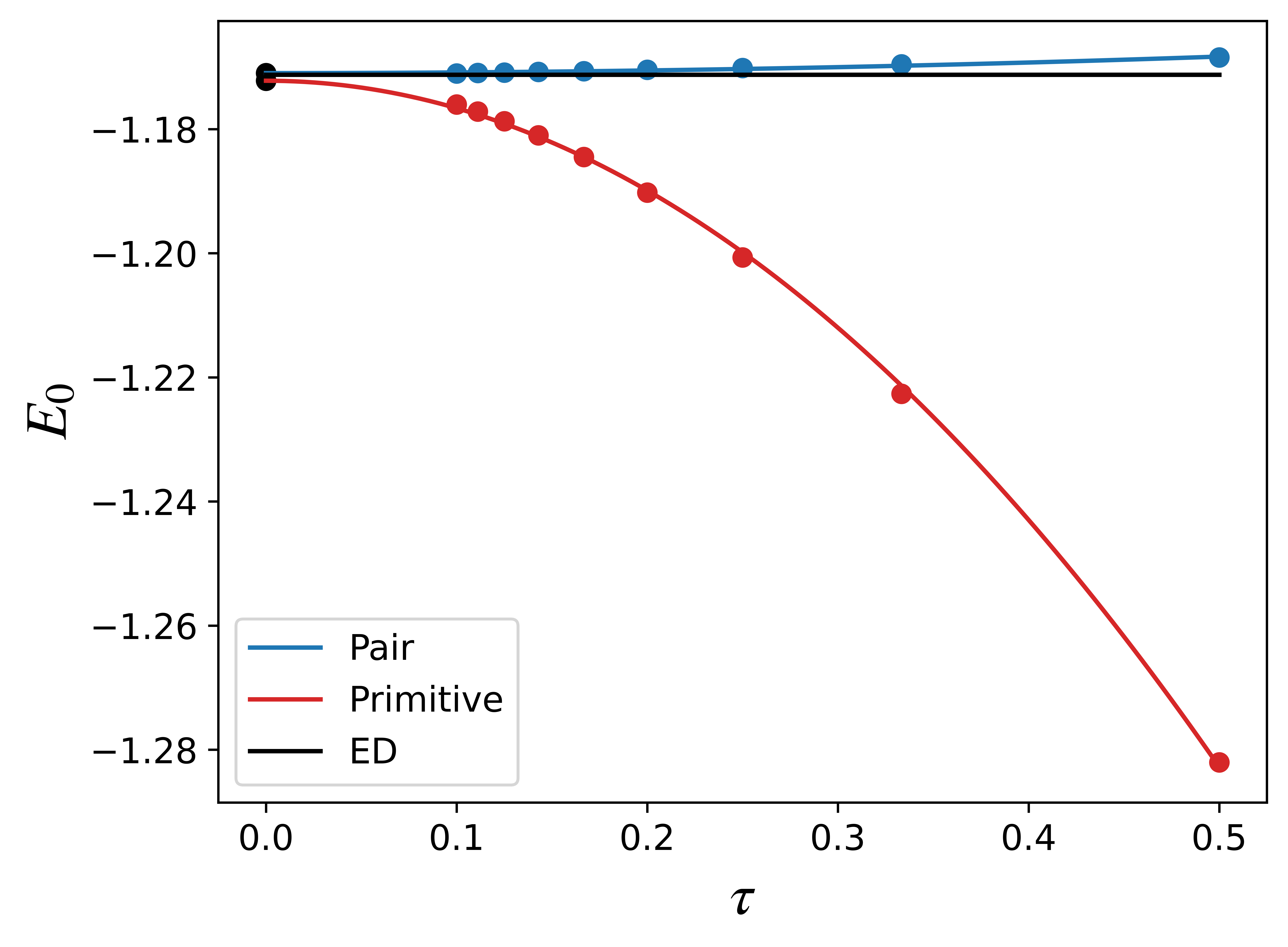}
\caption{Comparison of NMM ground state energy convergence in $\tau$ for the pair and primitive propagators for $N=3$. Here, $M=5$ and $g = 1.0$.} 
\label{NMM_Comparison}
\end{figure}

To benchmark the performance of the pair action against the symmetric Trotter factorization scheme, we first use NMM to compute the ground state energy for $3$ rotors. Figure \ref{NMM_Comparison} shows the results with $M=5$ and $g=1.0$. Exact diagonalization (ED) results are plotted for comparison. The primitive propagator here corresponds to the symmetric Trotter approximation and both extrapolations used quadratic fitting functions.  The pair factorization converges much faster than the typically used primitive factorization. As discussed above, this is because our potential is pair-wise additive which means that the $2$-body correlation effects are significant. These are accounted for exactly by the pair approximation. 

Practically, this implies that the pair approximation should yield a significant benefit in terms of efficiency when simulating many-body systems of rotating molecules with pair-wise interactions. This is particularly useful in Path integral Monte Carlo simulations where smaller $\tau$ values correspond to more degrees of freedom in the simulation. Therefore, improving convergence via better actions can greatly increase the efficiency of Monte Carlo simulations and allow us to computationally probe previously intractable regimes.\cite{ceperley_review} 

\begin{figure} [h]
  \centering
\includegraphics[width=\columnwidth]{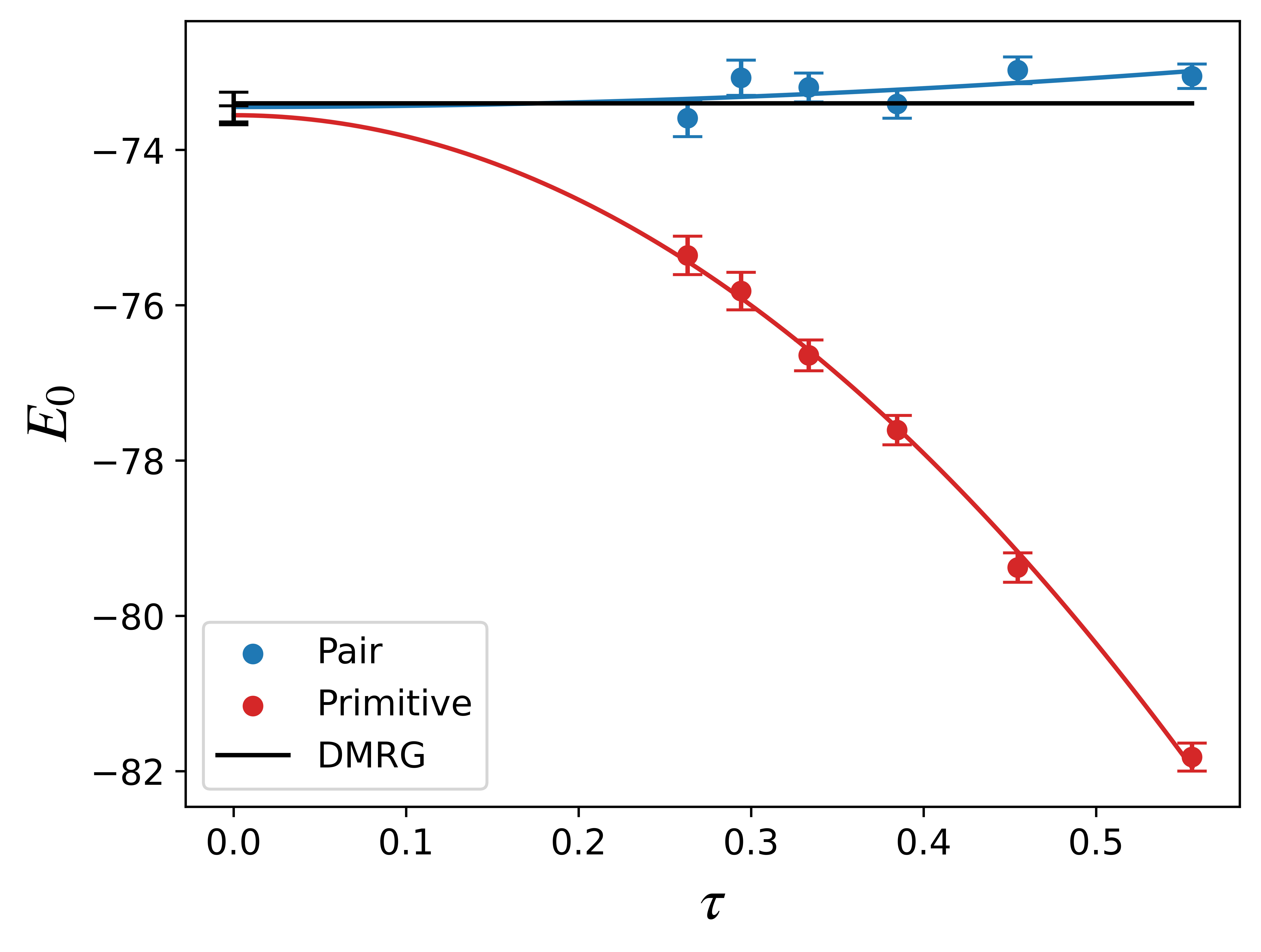}
\caption{Comparison of MC ground state energy convergence in $\tau$ for the pair and primitive propagators for $N=100$. Here, $M=5$ and $g = 1.0$. DMRG results are plotted to exhibit the exact value of the ground state energy.} 
\label{MC_Comparison}
\end{figure}
Figure \ref{MC_Comparison} exhibits the $\tau$ convergence comparison between the primitive and pair propagators for $N=100$. The ground state energy here was calculated using DVR-Gibbs Monte Carlo (MC) simulations. DMRG results are used to estimate the exact ground state energy for this case since exact diagonalization is unfeasible in the many-body regime. For these simulations as well, $g=1.0$ and $M=5$. Note that for larger systems, the primitive approximation would require more beads for convergence. However, increasing the number of beads would lead to simulations with smaller $\tau$ which, in the context of DVR-Gibbs could lead to the sign problem and exponential slowdown in convergence. Using a better action such as the pair propagator is very useful for such cases as well since it could then be used to attain convergence with a larger $\tau$, thus circumventing the sign problem. 
\subsection{Energetic Properties}
\begin{figure} [b]
  \centering
\includegraphics[width=\columnwidth]{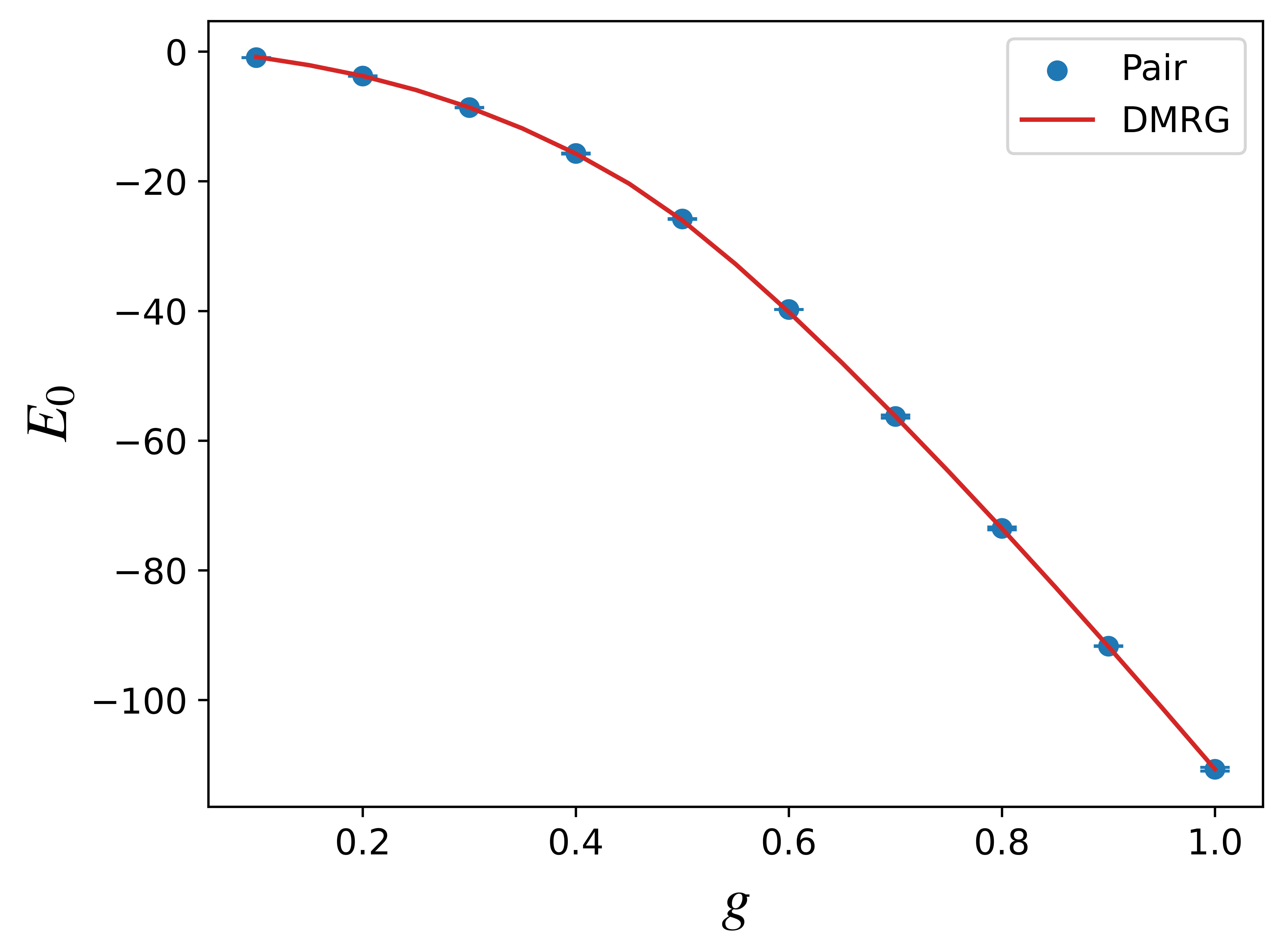}
\caption{Ground state energy as a function of $g$ for $N=150$ computed using the pair factorization compared to DMRG results.} 
\label{pair_energy}
\end{figure}

Figure \ref{pair_energy} shows the ground state energy computed using the pair propagator for a chain of $150$ planar rotors. For these results $M=10$ and $200000$ steps were used in the Monte Carlo simulation. All of the MC energy results were computed by extrapolation using quadratic fits with $P \in \{20,40,60\}$. The larger value of $M$ compared to the previous section was used to avoid the sign problem in the free DVR propagator for $P=60$. DMRG results are plotted for comparison which were calculated using $M=5$. Our method yields excellent agreement with DMRG simulations. 

\begin{figure} [!h]
  \centering
\includegraphics[width=\columnwidth]{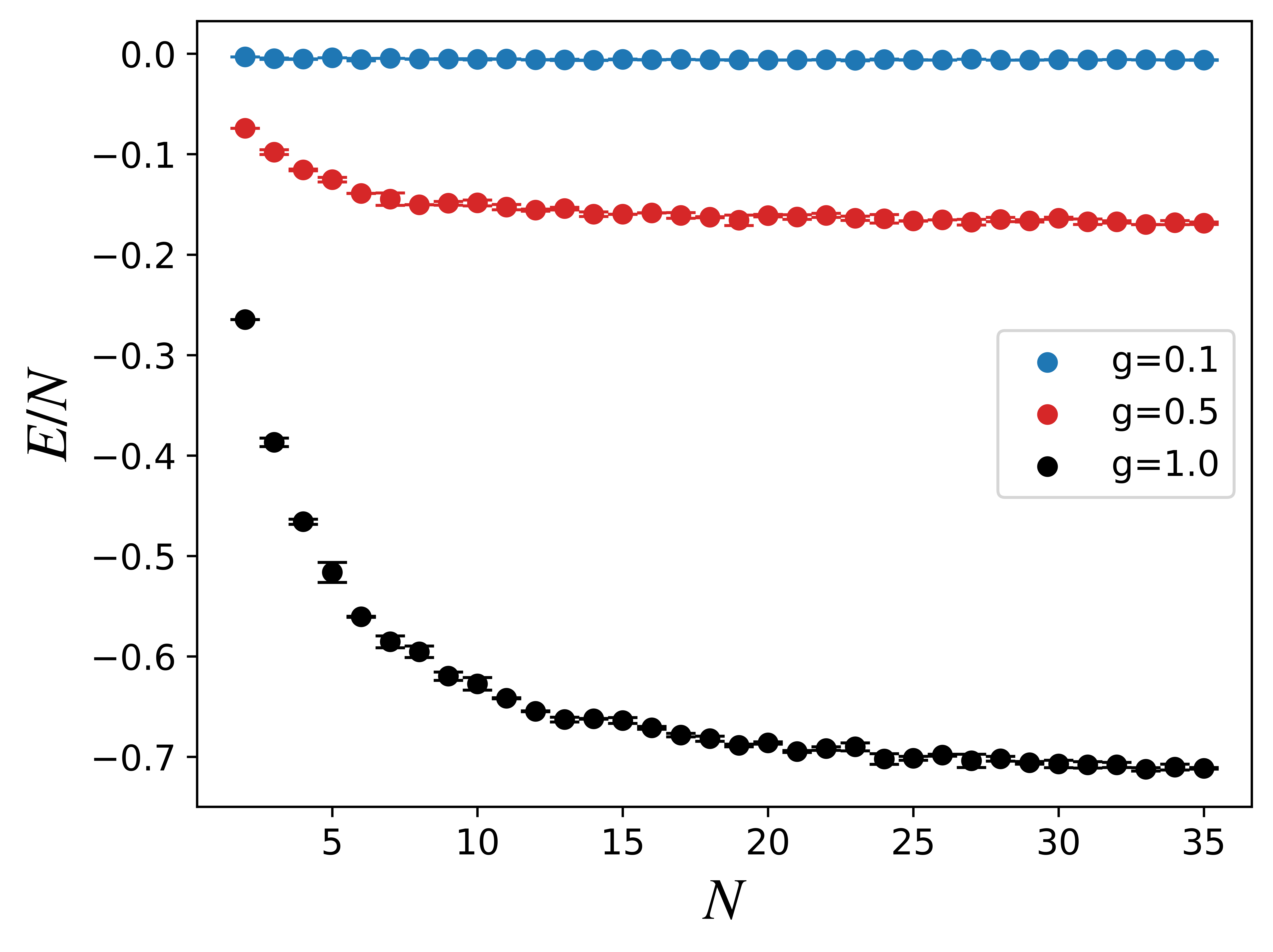}
\caption{Ground state energy per particle as a function of $N$ computed using the pair factorization approach for $g=0.1$, $g=0.5$ and $g=1.0$.} 
\label{chemical_potential}
\end{figure}

Figure \ref{chemical_potential} shows the energy par particle for this system computed using the pair propagator for $g \in \{0.1, 0.5, 1.0\}$. For $g=0.1$, the energy par particle converges very quickly due to the weak interactions. The $1/N$ behaviour of $E/N$ suggests a linear model for the ground state energy as a function of $N$, 
\begin{equation}
    E = \mu N + c
\end{equation}
where $\mu$ is the chemical potential. As $g$ is increased, larger values of $N$ are necessary for convergence to $\mu$. In particular for this system, we find that $N=35$ is sufficient in all three of these cases to see reasonable convergence in $E/N$. 
\subsection{Structural Properties} \label{struct_props}
\begin{figure*}
\includegraphics[width=.49\textwidth]{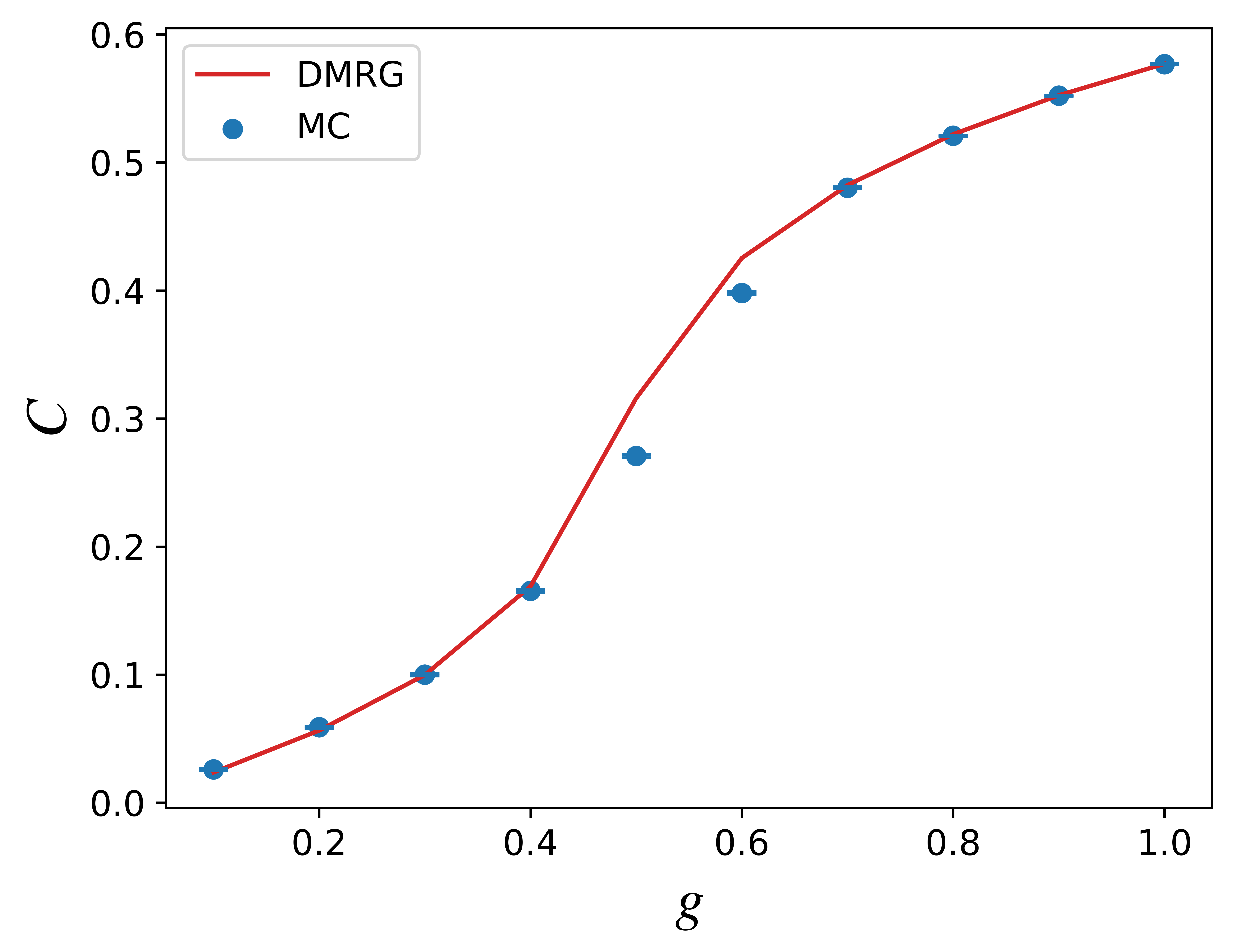}
\includegraphics[width=.49\textwidth]{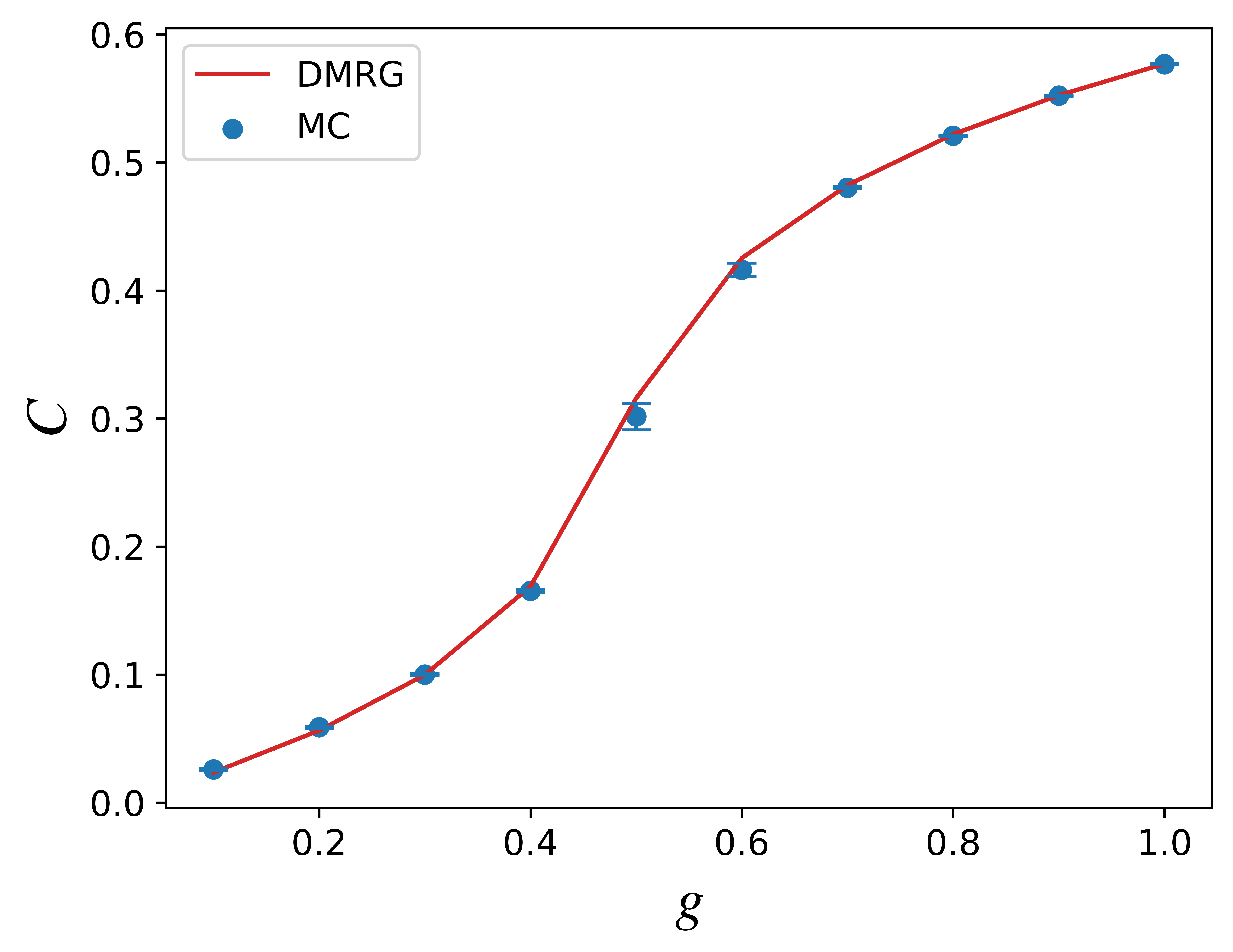}
\caption{Correlation as a function of $g$ for $N=150$ computed using the pair factorization compared to DMRG results. The left panel shows the MC results with $P=60$, $M=10$, and $3$ parallel simulations ($100000$ steps each) with different starting configurations. In the right panel, the point corresponding to $g=0.6$ is computed using a linear fit with $P \in \{50,60,70\}$, $M=12$ and $200000$ steps. For $g=0.5$, we again used a linear fit with $P \in \{80,90,100\}$, $M=14$ and $300000$ steps.} \label{pair_correlation_fig}
\end{figure*}

Figure \ref{pair_correlation_fig} shows the correlation as a function of coupling strength compared to DMRG. For $g < 0.5$, the system is in the disordered phase where the orientational correlation is close to $0$. For $g > 0.6$, the system is in the ordered phase which corresponds to higher correlations. For both of these regimes, we found that simulating the system with $60$ beads was sufficient for convergence to DMRG results. Three parallel simulation results were aggregated to generate the results shown in the left panel of figure \ref{pair_correlation_fig}; one intialized with all rotors oriented along the $+x$ axis, another with all rotors oriented along the $-x$ direction and a final one with all starting rotor orientations selected randomly using a uniform distribution. For each of these runs, we used $100000$ steps. The error bars in the left panel of figure \ref{pair_correlation_fig} correspond to the statistical errors from the Monte Carlo simulation. 

For $g=0.5$ and $g=0.6$, we needed to increase the number of beads to converge to the DMRG results. This is because  near criticality, many-body correlations become significant. The pair factorization only accounts for all $2$-body correlations exactly and as indicated by the error term in the series expansion (Eq. \eqref{P_3_error_term}), higher than $2$-body correlations determine the error in the pair propagated paths. When these become larger near critical points we need to decrease this systematic error by reducing $\tau$ further which can be achieved by increasing $P$. In particular for the orientational correlation in this system at $g=0.6$, we find that with $P \in \{50,60,70\}$, a linear fit yields the correct result. To circumvent the sign problem here, we increased $M$ to $12$. 

For $g=0.5$, we further needed to increase $P$ since we are closer to the transition point. Here, we find that with $P \in \{80,90,100\}$, a linear fit again yields the correct result when compared to DMRG. For this run, we used $M=14$ and kept track of the sign to account for the mild sign problem despite the larger $M$ value. These criticality corrected results are shown in the right panel of figure $\ref{pair_correlation_fig}$. The error bars in the right panel for all $g \notin \{0.5, 0.6\}$ correspond to statistical errors from the MC simulations whereas the error bars for $g=0.5$ and $g=0.6$ exhibit the fit errors. The number of simulation steps used for $g=0.6$ was $200000$ while for $g=0.5$, it was $300000$. We used a standard binning procedure \cite{mcmc_algo} to compute the statistical estimates for the orientational correlation throughout. We neglected the first $10000$ points in all of our correlation simulations to allow for the Markov chain to sufficiently equilibrate. 

\begin{figure}
  \centering
\includegraphics[width=\columnwidth]{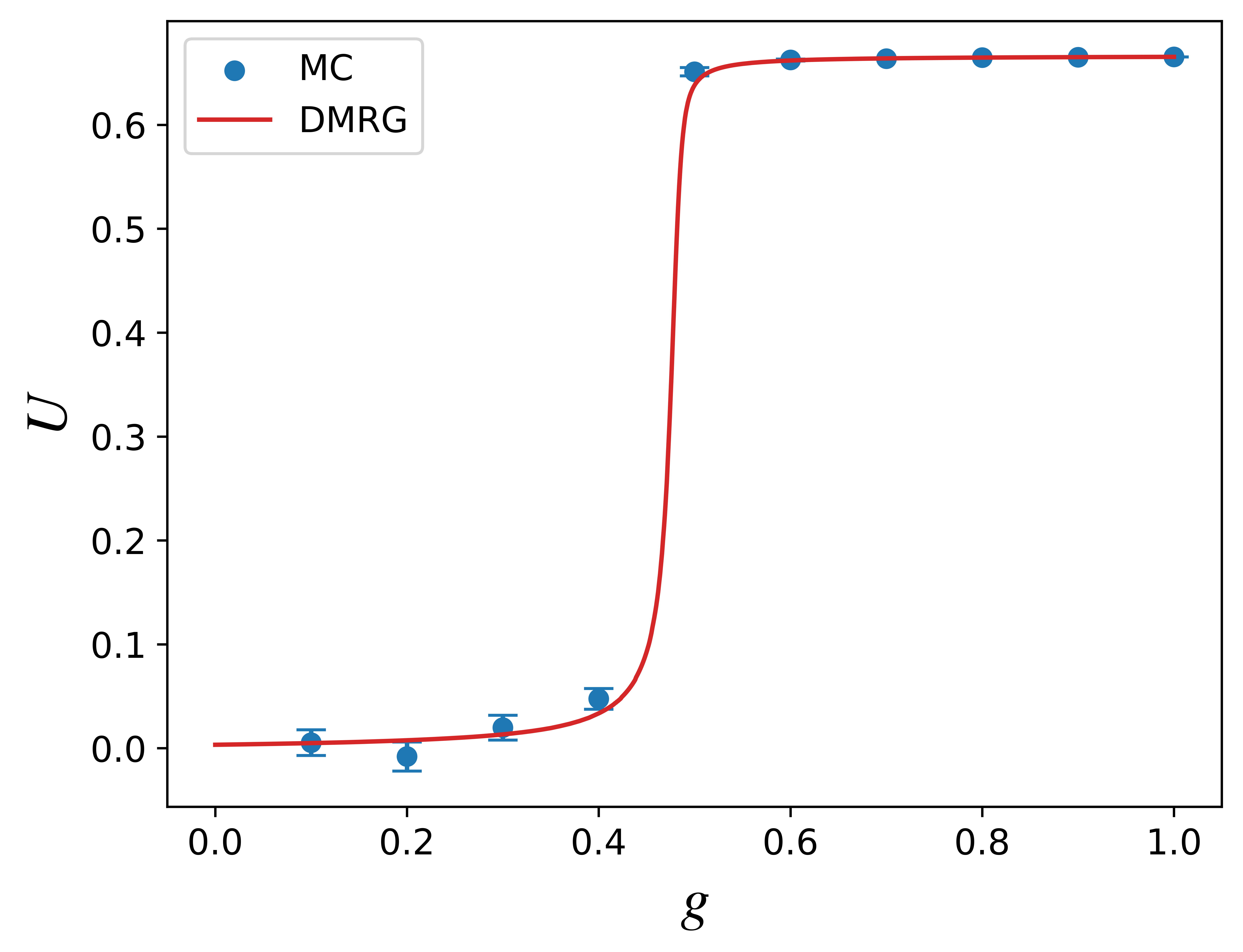}
\caption{Binder Ratio as a function of $g$ for $N=150$ computed using the pair factorization compared to DMRG results. For all values of $g$, the MC results were computed with $P=60$, $M=10$, and $100000$ steps. The system was initialized with all rotors pointing in the $+x$ direction.} 
\label{binder_fig}
\end{figure}

Figure \ref{binder_fig} shows the Binder ratio as a function of $g$ computed using the pair propagator compared to DMRG for $N=150$. For all of these points, the MC results were computed with $P=60$, $M=10$ and $100000$ steps. The system was initialized with all rotors along the $+x$ direction. We see good agreement with DMRG here as well and we find that we don't need a fit to converge to the exact result for the Binder ratio. However, the Binder ratio is more sensitive to the equilibration procedure and Markov chain auto-correlation near the phase transition. For $g < 0.5$, we use a burn-in period of $20000$ points for equilibration. For $g > 0.5$, we do not need to equilibrate at all since the starting configuration mimics the ground state configuration of the physical system. For $g=0.5$ and $g=0.6$, we are close to the critical region where the sytem is highly correlated as discussed before. Consequently, we need a much larger burn-in period for the chain to converge to the equilibrium distribution. For Fig. \ref{binder_fig}, we found that we needed to neglect the first $~99000$ points to see good agreement with the DMRG results. For the Binder ratio estimates, we used jack-knife analysis\cite{mcmc_algo} since that is more suitable for evaluating functions of random variables computed using correlated samples (we used the same simulation to compute $M_{x,0}^2$ and $M_{x,0}^4$ and the Binder ratio is a function of these two as specified in Eq. \eqref{Binder_Def}). We also neglected $5$ points between consecutive samples to ensure we were drawing independent samples from the Markov chain.   

\section{Conclusions} \label{conclusions_section}
In this work, we developed the pair-DVR PIMC approach to study the ground state properties of a linear chain of planar rotors with dipolar interactions. When it was initially introduced, the pair density was used in the context of Cartesian coordinates. We also analyzed the pair factorization path integral density error in powers of imaginary time using Wigner-Kirkwood expansions\cite{WK_expansion, WK_expansion2} and showed that the error is $\mathcal{O}(\tau^3)$ for our simulations. Computationally, we used the DVR-Gibbs sampling paradigm\cite{zhang2024path} to study the energetic and structural properties of our system as functions of $g$ and $N$. We also showed that the pair factorization converges much faster than the symmetric Trotter approximation in $\tau$. This is because for a pair-wise interaction, the equilibrium configuration is dominated by $2$-body correlations which are captured exactly by the pair propagator. Therefore, the sampled paths are much closer to the exact paths ($\tau \rightarrow 0$ limit) than the symmetric Trotter factorization, for a given $P$. We compared our results to DMRG and found excellent agreement for structural as well energetic properties in the large $N$ limit. This was done not only to showcase the utility of our approach in the context of improving our path integral simulation capabilities, but also to study the behavior of the pair approximation near criticality. 

This system undergoes a quantum phase transition at $g \approx 0.5$.\cite{planar_rotor_QPT} Therefore, long range correlations become important for converging to the correct equilibrium density in that regime. The pair action is unable to capture these many-body correlations exactly which means that we need more beads to converge to the correct results near $g \approx 0.5$. This is especially clear in our correlation results. Moreover, in our Binder ratio computations, we find that our sequential sampling techniques experience a critical slowing down\cite{Wolff_critical_slow_down} near $g \approx 0.5$ because our simulations require a large number of steps for equilibration. This underscores the necessity for adapting algorithms suitable for quantitatively studying phase transitions (for instance, the Wolff algorithm\cite{Wolff_critical_slow_down}) to rotational degrees of freedom with chemically relevant potentials such as those that employ dipolar and quadrupolar interactions. However, further research is needed to determine the utility of such approaches for the chain of rotors with dipolar interactions.

\section*{Acknowledgements}
The authors acknowledge the Natural Sciences and Engineering Research Council (NSERC) of Canada, the Ontario Ministry of Research and Innovation (MRI), the Canada Research Chair program, and the Canada Foundation for Innovation (CFI), and the Digital Research Alliance of Canada.

\section*{DATA AVAILABILITY}
The data that support the findings of this study are available from the corresponding author upon reasonable request.

\section*{References}
\bibliography{refs}

\clearpage

\renewcommand{\theequation}{S\arabic{equation}}
\setcounter{equation}{0}

\onecolumngrid
\appendix

Here, we provide some additional results pertaining to the pair propagator asymptotic error analysis that were omitted from the main text for the sake of brevity. 

\subsection{Green's Function Differential Equation} \label{supp_sect_a}
In the main text, we use the following differential equation for the quotient ($F(\vecphi, \vecphi', \tau)$) given by dividing the density matrix elements corresponding to the full Hamiltonian ($\rho(\vecphi, \vecphi', \tau)$) by the free propagator matrix elements (($\rho_0(\vecphi, \vecphi', \tau)$),
\begin{equation}
    \left(\frac{\partial}{\partial \tau} - \Delta_{\vecphi} - 2 \frac{\nabla_{\vecphi} \rho_0}{\rho_0} \cdot \nabla_{\vecphi} + V\right) F = 0, \ \ F = \rho/\rho_0 \label{green_diff_eq_supp}
\end{equation}
This can be derived using the Bloch equation for the free and the full propagators,  
\begin{gather}
    \left(\frac{\partial}{\partial \tau} + H\right) \rho(\vecphi, \vecphi', \tau) = 0, \ \ \lim_{\tau \rightarrow 0} \rho(\vecphi, \vecphi', \tau) = \delta(\vecphi - \vecphi') \label{full_H_bloch_supp}\\ \left(\frac{\partial}{\partial \tau} + K\right) \rho_0(\vecphi, \vecphi', \tau) = 0 
    , \ \ \lim_{\tau \rightarrow 0} \rho_0(\vecphi, \vecphi', \tau) = \delta(\vecphi - \vecphi') \label{free_H_bloch_supp}
\end{gather}
To see this, note that $H = K + V$ and $K = -\Delta_{\vecphi}$. Moreover, using our ansatz, we have that $\rho = F \rho_0$. Using this with Eq. \eqref{full_H_bloch_supp}, we get, 
\begin{equation}
    F \left(\frac{\partial \rho_0}{\partial \tau} - \Delta_\vecphi (\rho_0)\right) + \frac{\partial F}{\partial \tau} \rho_0 - \rho_0 \Delta_\vecphi (F) - 2 \nabla_{\vecphi} (\rho_0) \cdot \nabla_{\vecphi} (F) + V F \rho_0 = 0 \label{intermediate_1_supp}
\end{equation}
Eq. \eqref{free_H_bloch_supp} implies that the first term in the above equation is $0$. Dividing the resulting equation by $\rho_0$ yields Eq. \eqref{green_diff_eq_supp}. 

\subsection{Rotational Free Propagator Asymptotics} \label{supp_sect_b}
The many-body free propagator is a product of single particle free propagators which yields, 
\begin{gather}
    \rho_0 (\vecphi, \vecphi', \tau) = \prod_{i=1}^{N} \rho_0^{(1)}(\phi^i, \phi'^i, \tau) \nonumber
    \Rightarrow \left(\frac{\nabla_{\vecphi} \rho_0}{\rho_0}\right)_{j} = \frac{1}{\rho_0^{(1)} (\phi^j, \phi'^j, \tau)}\frac{\partial \rho_{0}^{(1)} }{\partial \phi_j} (\phi^j, \phi'^j, \tau)
\end{gather}
where $\rho_{0}^{(1)}$ is the single rotor free propagator. Using properties of Jacobi-Theta functions, it can be shown that, \cite{Whittaker_Watson_1996}  
\begin{equation}
     \left(\frac{\nabla_{\vecphi} \rho_0}{\rho_0}\right)_{j} = \sum_{n=1}^\infty (-1)^n \csch(\tau n) \sin(n (\phi^j-\phi'^j))
\end{equation}
Since we want an asymptotic expansion in $\tau$, we can use the series expansion of $\csch(n\tau)$ and retain the first non-trivial term to get:
\begin{equation}
    \left(\frac{\nabla_{\vecphi} \rho_0}{\rho_0}\right)_{j} \approx \frac{1}{\tau}\sum_{n=1}^{\infty} (-1)^n\frac{\sin(n (\phi^j-\phi'^j))}{n} 
\end{equation}
The above equation is just the Fourier series of the saw-tooth function divided by $\tau$ which, when restricted to $[-\pi, \pi)$ becomes: 
\begin{equation}
    \left(\frac{\nabla_{\vecphi} \rho_0}{\rho_0}\right)_{j} \approx -\frac{1}{2\tau} (\phi^j - \phi'^j) 
\end{equation}

\subsection{Second Term  Of The Wigner-Kirkwood Expansion} \label{supp_sect_c}
As discussed in the main text, Eq. \eqref{green_diff_eq_supp}, can be solved recursively using the Perelomov relation, 
\begin{gather} \label{Perelomov_recursion_supp}
    P_n(\vecphi, \vecphi') = n \int_0^1 dt \ t^{n-1} V(\vectheta(t)) P_{n-1}(\vectheta(t), \vecphi') - n \int_0^1 dt \ t^{n-1} \Delta_\vectheta P_{n-1}(\vectheta(t), \vecphi') 
\end{gather}
where $P_n$ is the $n$th term in the series expansion of $F$ in powers of $\tau$ ($F = \sum_{n} \frac{(-\tau)^n}{n!} P_n(\vecphi, \vecphi')$). Using Perelomov's recursion relation for $n=1$ and $n=2$ gives,  
\begin{gather}
    P_1(\vecphi, \vecphi') = \sum_{i=1}^{N-1}\int_{0}^{1} V_{i,i+1}(\vectheta(t)) \ dt = \sum_{i=1}^{N-1} W_{i, i+1} (\vecphi, \vecphi') \\
    P_2(\vecphi, \vecphi') = 2 \sum_{i, j=1}^{N-1}\int_0^1 dt \ t \ V_{i, i+1} W_{j, j+1}
    - 2\sum_{i=1}^{N-1} \int_0^1 dt \ t \left(\frac{\partial^2 W_{i, i+1}}{\partial \theta_i^2} + \frac{\partial^2 W_{i, i+1}}{\partial \theta_{i+1}^2}\right) \label{P_2_eqn_1_supp}
\end{gather}
where, $V_{i,i+1}=V_{i, i+1}(\vectheta(t))$, $W_{j,j+1}=W_{j,j+1}(\vectheta(t), \vecphi')$ in the above equation. We can simplify the above expression for $P_2$. To see this, define $I_{ij}$ to be a single term in the first summation of Eq. \eqref{P_2_eqn_1_supp}, 
\begin{equation}
    I_{ij} = \int_0^1 dt \ t \ V_{i, i+1}(\vectheta(t)) W_{j, j+1}(\vectheta(t), \vecphi') \label{I_i_j_eqn_supp}
\end{equation}
Using the definition of $W_{j,j+1}(\vectheta(t), \vecphi')$ where $\vectheta(t)$ represents a point on the line between $\vecphi$ and $\vecphi'$, we get, 
\begin{equation}
    W_{j,j+1}(\vectheta(t), \vecphi') = \frac{1}{t} \int_0^t V_{j,j+1}(\vectheta(t')) dt' \label{W_eqn_intermediate_supp}
\end{equation}
Using Eq. \eqref{W_eqn_intermediate_supp} and integrating Eq. \eqref{I_i_j_eqn_supp} by parts yields: 
\begin{gather}
    I_{ij} + I_{ji} = W_{i,i+1}W_{j,j+1}, \ \ I_{jj} = \frac{W_{j,j+1}^2}{2} \\ \Rightarrow 2\sum_{i,j=1}^{N-1} I_{ij} = 2\sum_{i<j}^{N-1} (I_{ij} + I_{ji}) + 2\sum_{j=1}^{N-1} I_{jj} = \sum_{j=1}^{N-1} W_{j,j+1}^2 + 2\sum_{i<j}^{N-1} W_{i,i+1}W_{j,j+1} \label{P_2_eqn_2_supp}
\end{gather} 
This gives the equation for $P_2$ provided in the main text, 
\begin{equation}
    P_2(\vecphi, \vecphi') = \sum_{j=1}^{N-1} W_{j,j+1}^2 + 2\sum_{i<j}^{N-1} W_{i,i+1}W_{j,j+1} 
    - 2\sum_{i=1}^{N-1} \int_0^1 dt \ t \left(\frac{\partial^2 W_{i, i+1}}{\partial \theta_i^2} + \frac{\partial^2 W_{i, i+1}}{\partial \theta_{i+1}^2}\right)
\end{equation}

\subsection{Pair Density Error} \label{supp_sect_d}

The diagonal terms ($\vecphi = \vecphi'$) of the Wigner-Kirkwood expansion are \cite{WK_expansion2}: 
\begin{gather}
    P_1 = V\\
    P_2 = V - \frac{1}{3}\Delta V\\
    P_3 = V^3 - V \Delta_{\vecphi} V - \frac{1}{2} (\nabla_\vecphi V)^2 + \frac{1}{10} \Delta_{\vecphi}^2 V \label{P_3_diag_eqn_supp}
\end{gather}
We can express $V^3$ from $P_3$ as follows: 
\begin{gather}
    V^3 = \sum_{n}V_{n,n+1}^3 + 6\sum_{n<m<l} V_{n,n+1}V_{m,m+1}V_{l,l+1} + 3 \sum_{n<m}(V_{n,n+1}^2V_{m,m+1} + V_{m_m+1}^2V_{n,n+1})
\end{gather}
$V\Delta_\vecphi V$ can be expressed as: 
\begin{gather}
    V \Delta_\vecphi V = \sum_{n} V_{n,n+1} (\Delta V)_{n,n+1} + \sum_{n<m} (V_{n,n+1} (\Delta V)_{m,m+1} + V_{m,m+1} (\Delta V)_{n,n+1}) \label{V_delta_V_supp}
\end{gather}
where we have used the notation: 
\begin{equation}
    (\Delta V)_{m,m+1} = \frac{\partial^2 V_{m,m+1}}{\partial \phi_m^2} + \frac{\partial ^2 V_{m,m+1}}{\partial \phi_{m+1}^2}
\end{equation}
Moreover, the bi-laplacian term $\Delta_{\vecphi}^2V$ is given by: 
\begin{gather}
    \Delta_{\vecphi}^2V = \sum_{m} (\Delta^2V)_{m,m+1} \\(\Delta^2V)_{m,m+1} = \frac{\partial^2 (\Delta V)_{m,m+1}}{\partial \phi_m^2} + \frac{\partial ^2 (\Delta V)_{m,m+1}}{\partial \phi_{m+1}^2}
\end{gather}
To compare $P_3(\vecphi, \vecphi)$ and $Q_3(\vecphi, \vecphi)$, first note that:
\begin{gather}
    Q_3 = \sum_{n=1}^{N-1} P_{3,n}^{(2)} + 6\sum_{n<m<l}P_{1,n}^{(2)}P_{1,m}^{(2)}P_{1,l}^{(2)} + \ 3\sum_{n<m}^{N-1} (P_{2,n}^{(2)}P_{1,m}^{(2)} + P_{1,n}^{(2)}P_{2,m}^{(2)}) \label{Q_3_eqn_supp}
\end{gather}
The first term in Eq. \eqref{Q_3_eqn_supp} can be expressed as, 
\begin{gather}
    \sum_{n}P_{3,n}^{(2)} = \sum_{n} (V_{n,n+1}^3 - V_{n,n+1} (\Delta V)_{n,n+1}) + \frac{1}{10}\sum_{n} (\Delta^2V)_{n,n+1} \nonumber \\ - \frac{1}{2}\sum_{n} \left( (\nabla V_{n,n+1})_{n}^2 +  (\nabla V_{n,n+1})_{n+1}^2\right)
\end{gather}
where $(\nabla V_{n,n+1})_{n}^2= (\partial_n V_{n,n+1})^2$ ($\partial_n$ here represents the derivative with respect to $\phi_n$). The second term in Eq. \eqref{Q_3_eqn_supp} can be written as, 
\begin{gather}
6\sum_{n<m<l}P_{1,n}^{(2)}P_{1,m}^{(2)}P_{1,l}^{(2)} = 6\sum_{n<m<l} V_{n,n+1}V_{m,m+1}V_{l,l+1} 
\end{gather}
The final term in Eq. \eqref{Q_3_eqn_supp} can be written as, 
\begin{gather}
3\sum_{n<m}^{N-1} (P_{2,n}^{(2)}P_{1,m}^{(2)} + P_{1,n}^{(2)}P_{2,m}^{(2)}) = 3 \sum_{n<m}(V_{n,n+1}^2V_{m,m+1} + V_{m_m+1}^2V_{n,n+1}) \nonumber \\ - \sum_{n<m} (V_{n,n+1} (\Delta V)_{m,m+1} + V_{m,m+1} (\Delta V)_{n,n+1})
\end{gather}
Using these, we find that $P_3$ and $Q_3$ agree for all terms in equation $\eqref{P_3_diag_eqn_supp}$ except for $-(\nabla_{\vecphi} V)^2/2$. The exact difference between $P_3$ and $Q_3$ is:
\begin{equation}
    P_3 - Q_3 = -\sum_{n=2}^{N-1} (\nabla V_{n-1,n})_{n} (\nabla V_{n,n+1})_{n} \neq 0 \label{P_3_error_term_supp}
\end{equation}
where we have used: 
\begin{gather}
    (\nabla_{\vecphi} V)^2 = \sum_{n=1}^{N-1} \left( (\nabla V_{n,n+1})_{n}^2 + (\nabla V_{n,n+1})_{n+1}^2\right) + 2 \sum_{n=2}^{N-1} (\nabla V_{n-1,n})_{n} (\nabla V_{n,n+1})_{n}
\end{gather}
Eq. \eqref{P_3_error_term_supp} is the error term provided in the main text.

\end{document}